# Magnetic hysteresis control in thin film Fe/Si multilayers by incorporation of $B_4C$

A. Zubayer[1], T. Hanashima[2], J. Sugiyama[2], A. David[4], X. Liu[4], N. Ghafoor[1], J. Stahn[3], A. Glavic[3], Y. Murakami[5], T. Takehiro[5], Y. Tomita[5], M. Auchi[5], J. Birch[1], M. Fahlman[4], F. Eriksson[1]

1. Department of Physics, IFM, Linköping University, SE-581 83 Linköping, Sweden
2. Neutron Science and Technology Center, Comprehensive Research Organization for Science and Society (CROSS), Tokai, Ibaraki 319-1106, Japan
3. PSI Center for Neutron and Muon Sciences, 5232 Villigen PSI, Switzerland
4. Department of Science and Technology, Laboratory of Organic Electronics (LOE), Linköping University, Norrköping, 60174, Sweden
5. The Ultramicroscopy Research Center, Kyushu University, Motooka 744, Nishi-ku, Fukuoka, Japan

## ABSTRACT

Magnetic hysteresis properties in Fe/Si multilayers have been studied as a function of the $B_4C$ content to control magnetization amplitude, coercivity, and hysteresis tilt, properties that are beneficial to tune for advancing applications in e.g. data storage, spintronics, and sensors. With an ion-assisted magnetron sputtering technique, 35 distinct thin film multilayer samples were prepared and their magnetic and structural properties were characterized by vibrating sample magnetometry, X-ray photoelectron spectroscopy, near edge X-ray absorption fine structure spectroscopy, and X-ray and neutron scattering methods. Key findings indicate that adding $B_4C$ lowers the coercivity and can decrease the saturation magnetization, demonstrating the tunability of magnetic responses based on composition. For samples with $\Lambda=30$Å periodicity, 10-15% of $B_4C$ addition produces antiferromagnetically (AF) coupled multilayers, and such AF coupling strength increases with the $B_4C$ content. Our findings reveal that B atoms do not chemically bind within the Fe atoms but instead occupy interstitial positions, disrupting medium- to long-range crystallinity thereby inducing the amorphization. Thereon, the observed effects on magnetic properties are directly attributed to this amorphization process caused by the presence of $B_4C$. The demonstrated ability to finely adjust magnetic properties by varying the $B_4C$ content offers a promising approach to overcome challenges in magnetic device performance and efficiency.

## INTRODUCTION

Measuring the magnetic hysteresis curve is a fundamental tool for analyzing the magnetic properties of materials, including thin films,[1–3] as it graphically represents the relationship between the applied magnetic field (H) and the induced magnetization (M). As H increases, M rises to saturation, but when H is reduced and reversed, M follows a different path. In other words, illustrating a hysteresis. The enclosed area of the hysteresis loop reflects magnetic energy losses due to internal friction and domain wall movement, quantified by the material's coercivity and remanence.[4] For thin films, these curves are invaluable in understanding microstructural behaviour, anisotropy, and interlayer exchange coupling, a quantum mechanical effect that governs the magnetic orientation between adjacent layers. Groundbreaking studies by physicists such as Weiss[5] and Stoner[6] have established the foundation for understanding magnetic domains and the influence of imperfections. The curves have been critical in exploring antiferromagnetic coupling between ferromagnetic layers separated by spacers,[7,8] leading to the discovery of the GMR effect, which earned Peter Grünberg the Nobel Prize in Physics[9] and has revolutionized data storage, spintronics, and applications relying on spin-dependent phenomena.[10] At low temperatures, magnetic properties like coercivity and magnetization are amplified as reduced thermal energy enhances magnetic moment alignment,[11] underscoring the delicate interplay of magnetic interactions in response to temperature and external fields.

In soft magnetic materials, such as those used in transformer cores, the hysteresis curve typically exhibits a narrow loop.[12] This narrowness indicates low coercivity and remanence, meaning the material can easily attain magnetization and demagnetization with minimal energy loss, which is ideal for applications involving frequent cycling of the magnetic field. For hard magnetic materials, like those used in permanent magnets, the hysteresis curve is much wider.[13] This indicates high coercivity and remanence, meaning the material maintains its magnetization even in the presence of substantial opposing magnetic field, crucial for permanent magnet applications. In polarizing neutron optics, which rely on materials that can polarize neutron beams via magnetic spin-dependent scattering, the external magnetic field, upon the optics operate within, needs to be large enough to saturate the optics.[14] These materials often have a high coercivity, but novel approaches incorporating $B_4C$ have shown to eliminate the coercivity to reach saturation for lower fields.[15] GMR devices, used in magnetic sensors and data storage technologies, exhibit hysteresis curves that are crucial for understanding the switching behaviour of the magnetic layers. These curves often display a tilted hysteresis indicating antiferromagnetic coupling, which is essential for the GMR effect.[16]

Controlling hysteresis in magnetic materials could enhance sensors and memory devices by allowing precise tuning over switching behaviours, improving a response time and reducing energy consumption. This capability would significantly optimize the performance of magnetic-based technologies. We investigate the potential to adjust the magnetization amplitude, coercivity, and the tilt of the hysteresis curve simply by integrating the material $B_4C$ into the Fe-based multilayer, Fe/Si. Additionally, we conducted a detailed investigation of the role of B in the Fe layers, revealing its unique impact on magnetic properties through the amorphization process, which alters the structural and magnetic behaviour of the Fe layers.

# RESULTS

The 35 samples were deposited on a Si substrate using an ion-assisted magnetron sputtering technique[17] (see Table 1). The nominal period thicknesses varied between 300 and 25 Å, while the number of periods varied between 2 and 24 so as to keep the total thickness at approximately 600 Å. Isotope enriched $^{11}B$ was used, instead of natural B, to increase the sensitivity for neutron measurements. The samples #1 to #32 all have the same amount of Fe and the $B_4C$ addition is increasing the multilayer volume and not replacing any amount of Fe or Si.

**Table 1. Samples.** All deposited samples with their respective number of periods (N), period thickness (Λ) and concentration of $B_4C$ throughout the multilayer in vol.%. The period thickness is solely nominal. The percentage $B_4C$ is an added value, thus the period is in actuality increasing with $B_4C$ content. The layer thickness ratio Fe/(Fe+Si) was aimed to be 0.5, thus equally thick Fe as Si layers.

| Sample # | Periods (N) | Period thickness ($\Lambda$) [Å] | Percentage $B_4C$ [%] |
|---|---|---|---|
| Sample 1 | 2 | 300 | 0 |
| Sample 2 | 2 | 300 | 2.5 |
| Sample 3 | 2 | 300 | 5 |
| Sample 4 | 2 | 300 | 10 |
| Sample 5 | 2 | 300 | 15 |
| Sample 6 | 2 | 300 | 20 |
| Sample 7 | 2 | 300 | 40 |
| Sample 8 | 2 | 300 | 80 |
| Sample 9 | 6 | 100 | 0 |
| Sample 10 | 6 | 100 | 2.5 |
| Sample 11 | 6 | 100 | 5 |
| Sample 12 | 6 | 100 | 10 |
| Sample 13 | 6 | 100 | 15 |
| Sample 14 | 12 | 50 | 0 |
| Sample 15 | 12 | 50 | 2.5 |
| Sample 16 | 12 | 50 | 5 |
| Sample 17 | 12 | 50 | 10 |
| Sample 18 | 12 | 50 | 15 |
| Sample 19 | 20 | 30 | 0 |
| Sample 20 | 20 | 30 | 2.5 |
| Sample 21 | 20 | 30 | 5 |
| Sample 22 | 20 | 30 | 7.5 |
| Sample 23 | 20 | 30 | 10 |
| Sample 24 | 20 | 30 | 12.5 |
| Sample 25 | 20 | 30 | 15 |
| Sample 26 | 24 | 25 | 0 |
| Sample 27 | 24 | 25 | 2.5 |
| Sample 28 | 24 | 25 | 5 |
| Sample 29 | 24 | 25 | 7.5 |
| Sample 30 | 24 | 25 | 10 |
| Sample 31 | 24 | 25 | 12.5 |
| Sample 32 | 24 | 25 | 15 |
| Sample 33 | 4 | 500 | 0 |
| Sample 34 | 4 | 500 | 15 |
| Sample 35 | 4 | 500 | 20 |

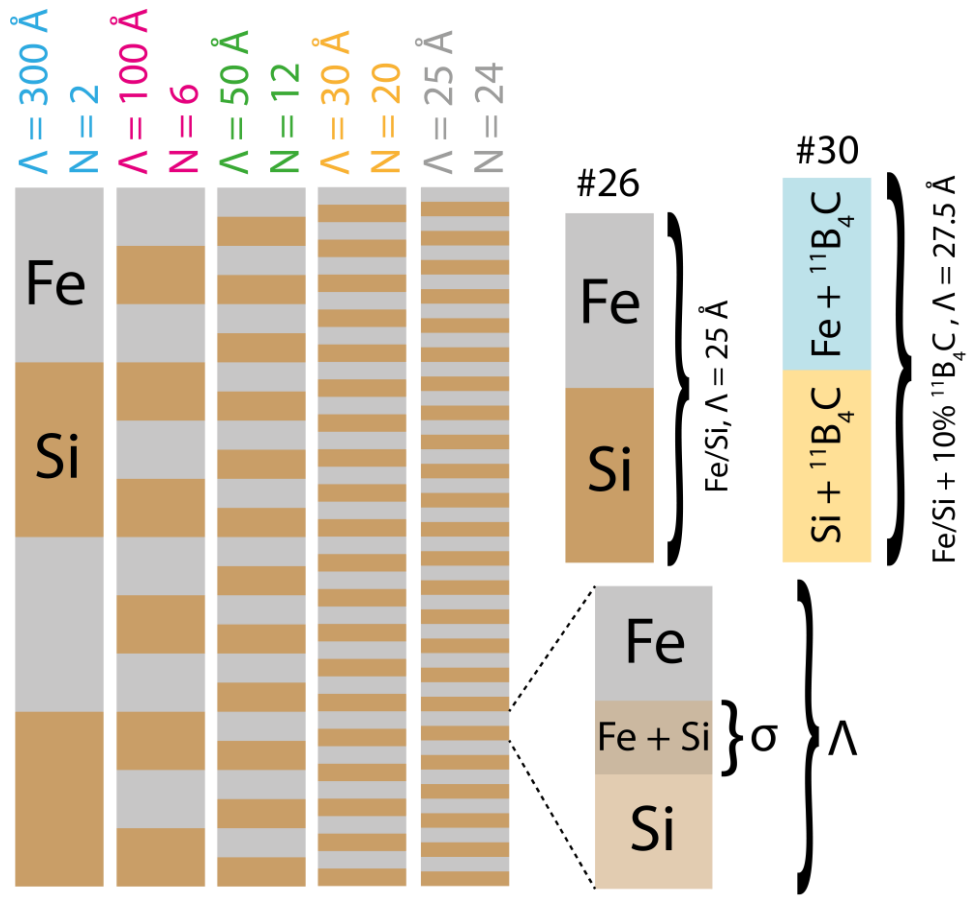

**Figure 1. Schematics.** Sketch of the sample variations seen in Table 1 (#1-32). Included are illustrations that although samples #26 and #30 belong to the same category $\Lambda = 25$ Å, the B$_4$C samples uses an addition of B$_4$C and not a substitution, thus maintaining the same amount of Fe atoms but slightly increasing the $\Lambda$. The same applies for all categories. Further, an illustration of what period thickness $\Lambda$ and interface width $\sigma$ is. Each interface of all multilayers have interfaces, where the higher the number of periods, N, the higher the number of interfaces as well.

Figure 1 illustrates the multilayers grown as well as informatively describes the nomenclature. Further that a higher N will have more interfaces resulting in a higher amount of Fe and Si mixed regions, meaning that the $\Lambda = 25$ Å series will have the highest amount of Fe + Si mixed material in the sample.

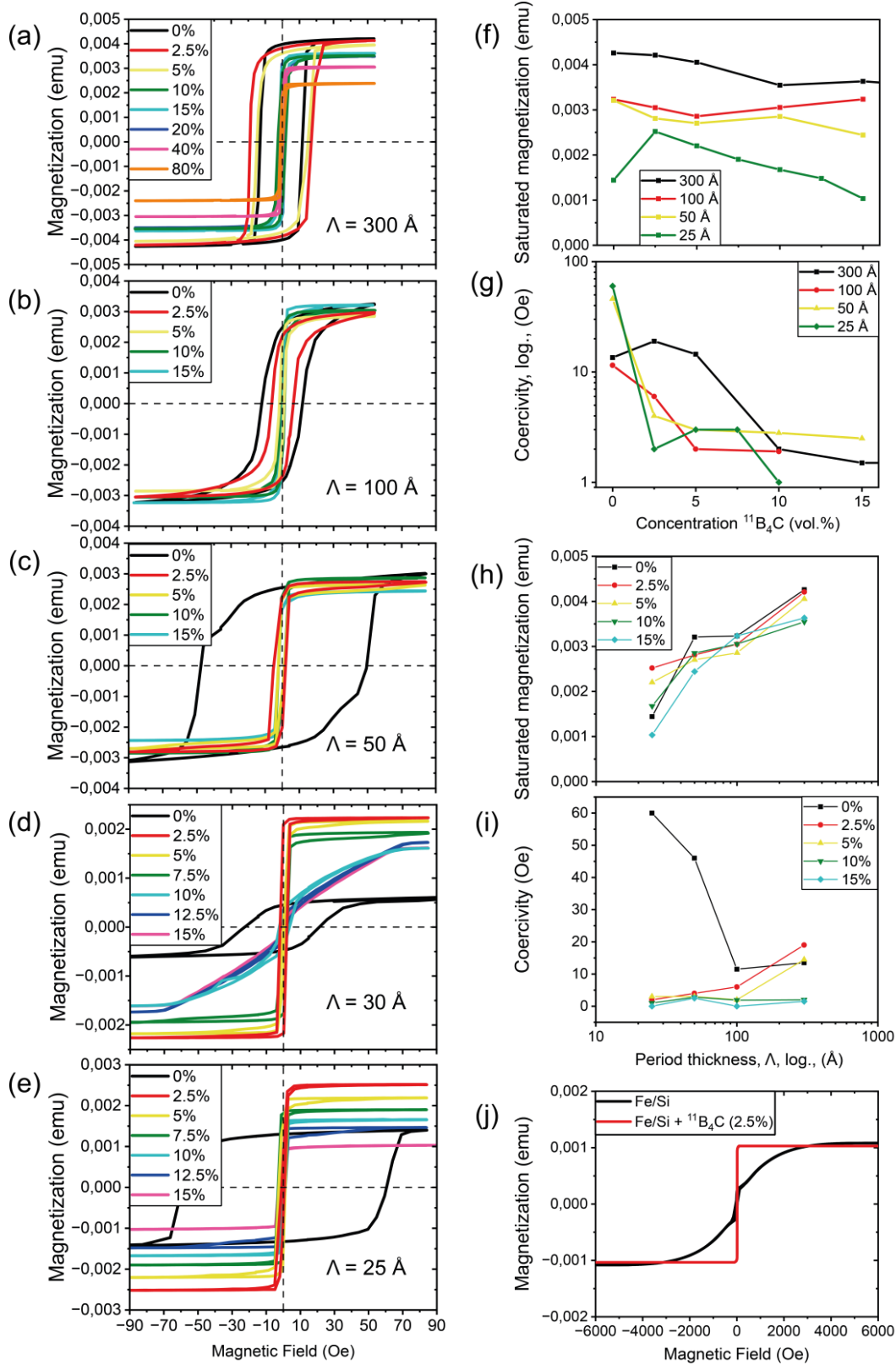

**Figure 2. Magnetometry.** Vibrating sample magnetometry on 25 samples (Sample #1–32). (a-e) shows the magnetic hysteresis curves for the series of Λ = 300, 100, 50, 30 and 25 Å, respectively. (f) Saturated magnetization as a function of concentration. g) Coercivity as a function of concentration. (h)Saturated magnetization as a function of period thickness going from large period down to small period. (i) Coercivity as a function of period thickness going from large period towards smaller periods. (j) showing Fe/Si + B$_4$C with 0 and 2.5 % from (d) in a field range between -6000 and 6000 Oe.

All the samples were measured using vibrating sample magnetometry (VSM) to obtain the hysteresis curves as seen in Figure 2(a-e), where (a-e) shows the hysteresis for each period thickness 300, 100, 50, 30 and 25 Å respectively. All samples measured showed considerable magnetization, even when 80 % of the period of the Fe/Si multilayer consisted of $B_4C$. It is apparent that pure Fe/Si multilayers always have a large magnetic coercivity, while the addition of $B_4C$ decreases this coercivity until the coercivity disappears. For the samples with Λ=300 Å and 25 Å, it is clear how the increasing $B_4C$ content decreases the saturated magnetization as well, besides when comparing the pure Fe/Si 25 Å with 2.5-10 % $B_4C$ Fe/Si 25 Å. For the samples with Λ=100 Å and 50 Å, the addition of $B_4C$ does not alter the saturated magnetization. Note that the saturated magnetization values seen in Figure 2(f) is obtained from the magnetization at H=100 Oe, where almost all the samples are arguably saturated. This subplot shows that M decreases as a function of x. Figure 2(g) shows the coercivity as a function of x for each period thickness. This indicates a decrease in coercivity with increasing x, besides the samples with Λ=300 Å and x=2.5 vol.%. Figure 2(h) shows the saturated magnetization as a function of Λ in log scale. The magnetization always decreases with increasing Λ, regardless of x. Figure 2(i) shows the coercivity as a function of Λ for x=0, 2.5, 5, 10 and 15 vol.%. The coercivity is always negligeable when x ≥ 10%, while the coercivity clearly decreases for thinner Λ with x=2.5 and 5%. However, for pure Fe/Si, the coercivity increases significantly for Λ=50 and 25 Å. The Fe/Si multilayer seems to not be fully saturated even at 100 Oe as seen in Figure 2(d), but as seen in 2(j) the Fe/Si sample is not saturated until 5000 Oe is reached.

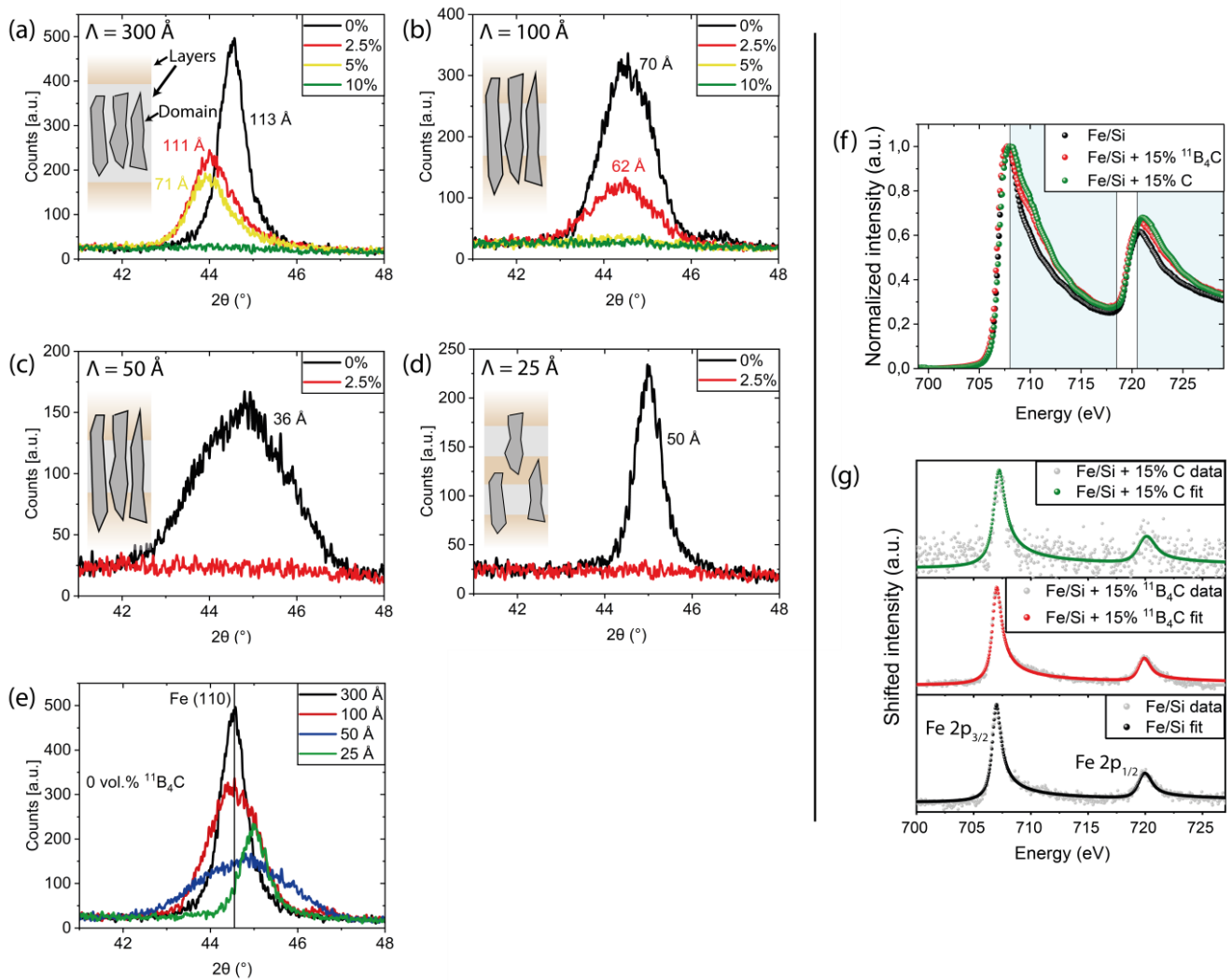

**Figure 3. X-ray diffraction and spectroscopy.** (a-e) X-ray diffraction on 12 samples. (a) on samples #1-4, (b) on samples #9-12, (c) on samples #14-15, and (d) on samples #26-27. (e) on samples #1, #9, #14 and #26. The calculated out-of-plane crystallite domain sizes are written for each sample if the Fe (110) peak was visible. This was roughly calculated using the Scherrer equation. Further for samples containing any size of crystallites a schematic is shown depicting the domain size in contrast to layer size, indicating if the out-of-plane crystallite size are larger or smaller than the Fe layer thickness for each case. (f-g) Near-edge X-ray absorption fine structure spectroscopy around the Fe $L_{2,3}$ absorption edge (a) and X-ray photoelectron spectroscopy (b) of Fe/Si (#19), Fe/Si + 15% $B_4C$ (#25) and a reference Fe/Si + 15% C sample ($\Lambda$ = 100, N = 10). In (a) the highlighted blue background area indicates the difference from the Fe/Si sample. (b) shows the XPS focusing on the two Fe peaks.

The XRD patterns around the Fe peak are shown in Figure 3. Figure 3(a) shows the XRD patterns for the samples #1-4 with the periodicity of 300 Å. The XRD pattern for the 10% $B_4C$ sample lacks a diffraction peak, indicating that the approximate requirement to amorphize Fe layers with a thickness of approximately 150 Å is somewhere between 5 and 10%. In this section amorphous means X-ray amorphous, since in order to see a diffraction peak the lattice planes must be coherent over a distance larger than the coherence length. For the samples with the $B_4C$ content between 0 – 5%, the height of the diffraction peak decreases with increasing $B_4C$ content. On the other hand, the peak position shifts towards lower angles by about 0.5° with the $B_4C$ introduction. As seen in Figure 3(b), for the 100 Å periodicity samples, only 5% $B_4C$ introduction is enough to amorphize the Fe layer, judging from the absence of the diffraction peak in the 5% sample. As seen in Figure 3(c), for the 50 Å periodicity samples, only 2.5% $B_4C$ introduction leads to the amorphous state of the Fe layer. The amorphous phase is also clearly observed for the 2.5% $B_4C$ in Fe/Si multilayers sample with a periodicity of 25 Å [Figure 3(d)]. What is notable from the crystallite sizes is that they can have an out-of-plane size larger than the Fe layer thickness, indicating a large interface width due to crystallites. The Fe/Si sample with $\Lambda$ = 25 Å shows a crystallite size equal to an entire period, indicating heavily mixed Fe and Si layers. It is important to note that the Scherrer equation provides a very approximate estimation of crystallite size, as it does not account for factors such as strain, instrumental broadening, or anisotropic effects, which can significantly influence the broadening of diffraction peaks. When $B_4C$ is incorporated the crystallite size initially decreases but then turns X-ray amorphous with increasing $B_4C$ concentrations. Figure 3(e) shows the XRD patterns for the pure Fe/Si multilayers with the 4 different periods 300, 100, 50 and 25 Å. The height of the diffraction peak decreases from decreasing the period from 300 to 50 Å, while the height for the diffraction peak in the 50 Å sample is lower than that in 25 Å, but with a notable change in 2θ to higher angles.

A way to selectively probe the elements in these systems is to use NEXAFS, which is a more bulk-sensitive technique by collecting drain current.[18] In this case, the Fe $L_{2,3}$ absorption edge was measured to probe the resonant transition from 2p to empty 3d states. The Fe 2p NEXAFS in three different samples demonstrated nearly identical spectra indicating homogenous sample compositions, as shown in Figure 3(f). The line shapes are very similar for sample Fe/Si + $B_4C$ and Fe/Si + C, both present a shoulder at the $L_3$ edge around 710 eV. However, for sample Fe/Si the shoulder feature is rather weaker at the $L_3$ edge. In the case of sample Fe/Si the $L_{2,3}$ branching ratio ($I(L_3)/[I(L_2)+I(L_3)]$) is 0.62 which is slightly higher than the 0.60 of the other samples. Chemical bond formation between Fe and B is possible but generally occurs at higher temperatures than the deposition temperature here.[19] Compared to the reference sample containing Fe + C, Fe/Si + 15% $B_4C$ (#25) sample shows a similar shoulder making it the most likely to assign the shoulder formation to Fe-C bond formation.

To investigate the chemical state and potential chemical bonds, X-ray photoelectron spectroscopy (XPS) measurements were employed with a focus on Fe 2p and Si 2p regions. Figure 3(g) shows XPS Fe 2p results for all three samples (after U3 Tougaard background subtraction which is counteracting the inelastic scattering effects present in this layered structure). All Fe 2p spectra show very similar spectral feature to that in the pure Fe metal, which indicates the chemical state is likely with Fe-Fe bonds, while other bonds are negligeable.[20] Thus, the amorphization does not seem to be due to Fe-B bonds.

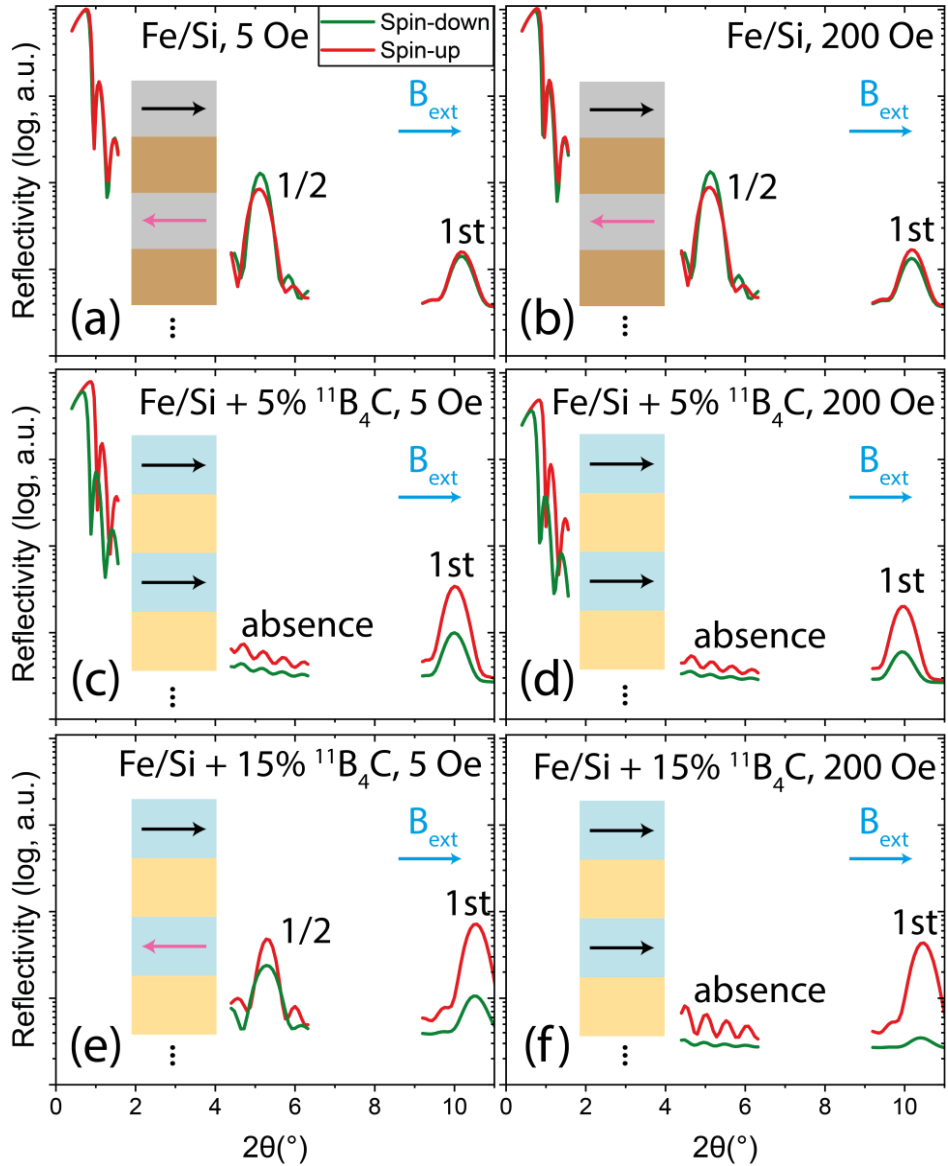

**Figure 4. Neutron scattering.** Polarized neutron reflectivity (PNR) fits from measurements on 3 samples (Sample #26, #28 and #32), which all have a periodicity of $\Lambda=30$ Å, N = 20, and with the concentration of $B_4C$ being 0 (a-b), 5 (c-d)) and 15 (e-f) vol.%. (a, c and e) are PNR plots when the external field was 5 Oe, while (b, d and f) was measured for 200 Oe. The plots focus only on the critical edge region, half-order Bragg peak region and the first-order Bragg peak region. Schematics of the magnetization alignment in each magnetic layer corresponding to the measurements are shown in every inset.

Figure 4 shows the result of polarized neutron reflectivity (PNR) fits from measurements of the 3 samples: i.e., #26, 28 and 32, which all belong to the 30 Å periodicity series. Figures 4(a) and 4(b) show the PNR spectra for the pure Fe/Si sample with two different external field: that is, 5 and 200 Oe. The half-order Bragg peak is prominent regardless of these specific external fields. On the contrary, reflectivity of the first-order Bragg peak seems to increase slightly for the spin-up neutrons case, while decrease for the spin-down case, when the external field is 200 Oe. Figures 4(c) and 4(d) show the PNR of 5% $B_4C$ in Fe/Si, which have no half-order Bragg peak but a first-order Bragg peak for both 5 and 200 Oe. Finally, Figures 4(e) and 4(f) show the 15% $B_4C$ in Fe/Si which has a half-order Bragg and a first-order Bragg peak with low polarization for 5

Oe external field. While at 200 Oe the half-order Bragg peak is eliminated while the first Bragg peak is significantly better polarized.

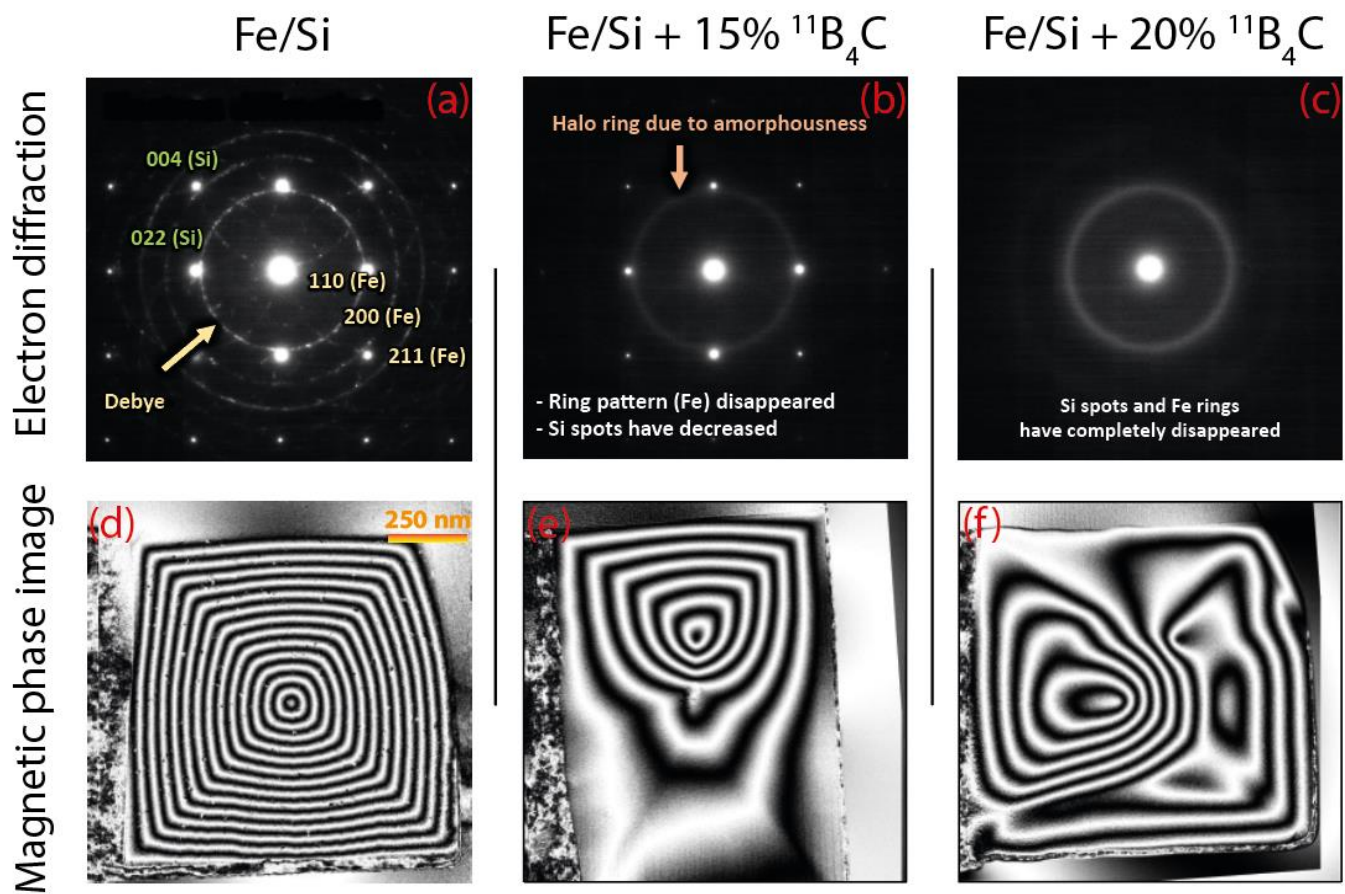

**Figure 5. Electron microscopy.** Electron diffraction (a-c) and magnetic phase images (d-f) in electron holography of Fe/Si (#33), Fe/Si + 15% $B_4C$ (#34) and Fe/Si + 20% $B_4C$ (#35), respectively.

The results of electron holography as seen in Figure 5, is a method utilizing TEM, which revealed significant differences in the magnetic domain structures across the crystalline (Fe/Si), almost amorphous (Fe/Si + 15% $B_4C$), and fully amorphous (Fe/Si + 20% $B_4C$) samples, when there is no external field component across the surface (no component in-plane). Electron holography provides the contour lines, as show in Figure 5, in terms of the phase shift of the incident electron wave.[21] These contour lines approximate the magnetic flux lines projected in the foil plane. Spacing of the contour lines is inversely proportional to the magnetic flux density multiplied by the foil thickness. In the pure Fe/Si crystalline sample, thinner and well-defined contour lines indicate uniform and coherent in-plane magnetic domains, reflecting strong interlayer magnetic coupling and structural order. In contrast, the almost amorphous sample, Fe/Si + 15% $B_4C$, exhibits thicker and more meandering contour lines within the foil. This suggests reduced interlayer magnetic interactions and local variations in anisotropy due to partial amorphization. The fully amorphous sample, Fe/Si + 20% $B_4C$ shows highly complex magnetic contour distributions, with overlapping contributions from different magnetic domain structures across the layers.

# DISCUSSION

## Control of magnetic hysteresis loop

The results in Figure 2 show the saturated magnetization amplitude for samples with varying amounts of $B_4C$. The magnetization amplitude in a material is governed by the ferromagnetic coupling between the neighboring atoms: specifically, the greater the number of neighboring magnetic atoms, the higher the magnetization.[22] When the material is diluted with $B_4C$, although the total number of Fe atoms remains constant, the reduced number of neighboring Fe atoms due to the presence of non-ferromagnetic atoms leads to a decrease in magnetization amplitude.[15] Such a dilution effect vs magnetization process allows us to tune the magnetization in X/Y multilayers depending on the $B_4C$ concentration, where X is a ferromagnetic material such as Fe and Y is a non-magnetic material such as Si. However, for the 100, 50 and 25 Å periodicity, the decrease in magnetization is not observed in the field range within hundreds of Oe, due to a weak antiferromagnetic coupling for pure Fe/Si multilayers, which is common for Fe layers between 15 and 50 Å.[23] The antiferromagnetic coupling between Fe layers naturally requires higher external fields to align all Fe layers' magnetization parallel. Note that a tilt in a hysteresis curves is typical for the multilayers with an antiferromagnetic coupling [Figures 2(b, c and e)].

The coercivity can also be controlled by incorporating specific amounts of $B_4C$. The coercivity is related to the crystallinity of the ferromagnetic layers.[24] The higher the crystallinity of the Fe layer the larger the coercivity. The incorporation of $B_4C$ leads to amorphization of the Fe layers, as also evident in Figure 3. Comparing the intensity of the Fe diffraction peak in the XRD patterns [Figures 3(a-e)] as a function of coercivity [Figures 2(a, b, c, e) or 2(f)], it is clear that, as the Fe diffraction peak has perished in the XRD pattern, the coercivity also diminishes. When such diffraction peak is weakened due to an introduction of a few percent of $B_4C$, the coercivity is weakened as well. Thus, the required concentration of $B_4C$ for the desired outcome can be extracted. Therefore, the optimal concentration ranges between the fully amorphous concentration and 0% so as to achieve the desired coercivity. Note however that the thinner the Fe layers are, the smaller amount of $B_4C$ is required to eliminate the coercivity.

Several linear, tilted, M-H curves are seen in Figure 2(d) for the samples with a periodicity of 30 Å and $B_4C$ concentrations 10% and above. This tilted behavior is common for the antiferromagnetic materials or antiferromagnetically coupled ferromagnetic layers.[25] Another possibility could also be due to magnetic domains within the layers which gives rise to tilted M-H curves.[26] The pure Fe/Si samples also exhibit a slight tilted increase in M with H even beyond its coercive fields. As seen in Figure 2(j) the Fe/Si sample is not saturated until approximately 5000 Oe is reached. To investigate such tilted behaviors, the PNR results (Figure 4) is coupled with the VSM results. Half-order Bragg peaks would only appear if there is a second periodicity, such as an antiferromagnetic coupling between Fe layers causing a magnetic variation with a period twice as large as a bilayer. The pure Fe/Si sample had a half-order Bragg peak for both 5 and 200 Oe meaning that the antiferromagnetic coupling is persistent enough to not let the Fe layers saturate at 200 Oe. The first-order Bragg peak is more polarized for 200 Oe than 5 Oe, which is expected due to the Fe/Si being the common polarizer material system for neutron optics, where the polarization is better the closer to saturation reached. The 5% sample showed no tilt behavior in the hysteresis seen in Figure 2, and also had no half-order Bragg peak at 5 or 200 Oe in PNR, as expected. The first-order Bragg peak and its polarization was well defined and prominent for 200 Oe, in agreement with literature. Hence the 5% sample has no antiferromagnetic coupling. The 15% sample has the linearly tilted behavior in the hysteresis in Figure 2(d) and an antiferromagnetic half-order Bragg peak at 5 Oe. Further, as expected, the half-order Bragg peak is gone at 200 Oe in good agreement with the hysteresis curve in being fully saturated already at 70 Oe. In other words, the 15% sample is antiferromagnetically ordered only within the external field range of -70 Oe to 70 Oe, since the external magnetic force overpowers the antiferromagnetic coupling outside this field range. The reason as to why only a certain period thickness indicate antiferromagnetic coupling is due to a critical spacer thickness that optimizes interlayer exchange interactions.[7,23] It seems as if the addition of $B_4C$ further enhances AFM coupling at 30 Å, potentially by promoting amorphization, reducing crystallinity, and improving spacer uniformity. This behaviour contrasts with larger periods (50 Å, 100 Å, 300 Å), where weaker coupling arises from reduced overlap of exchange fields. For smaller periods (25 Å), the system remains ferromagnetic (FM) but may no longer exhibit effective coupling, as interlayer mixing and structural disruptions dominate,[27] and the spacer layer approaches an insulating-like state,[28] limiting its ability to mediate interactions. These results emphasize the pivotal role of spacer thickness and composition in modulating magnetic properties, underscoring how the interplay between structural and electronic factors shapes the observed behaviour.

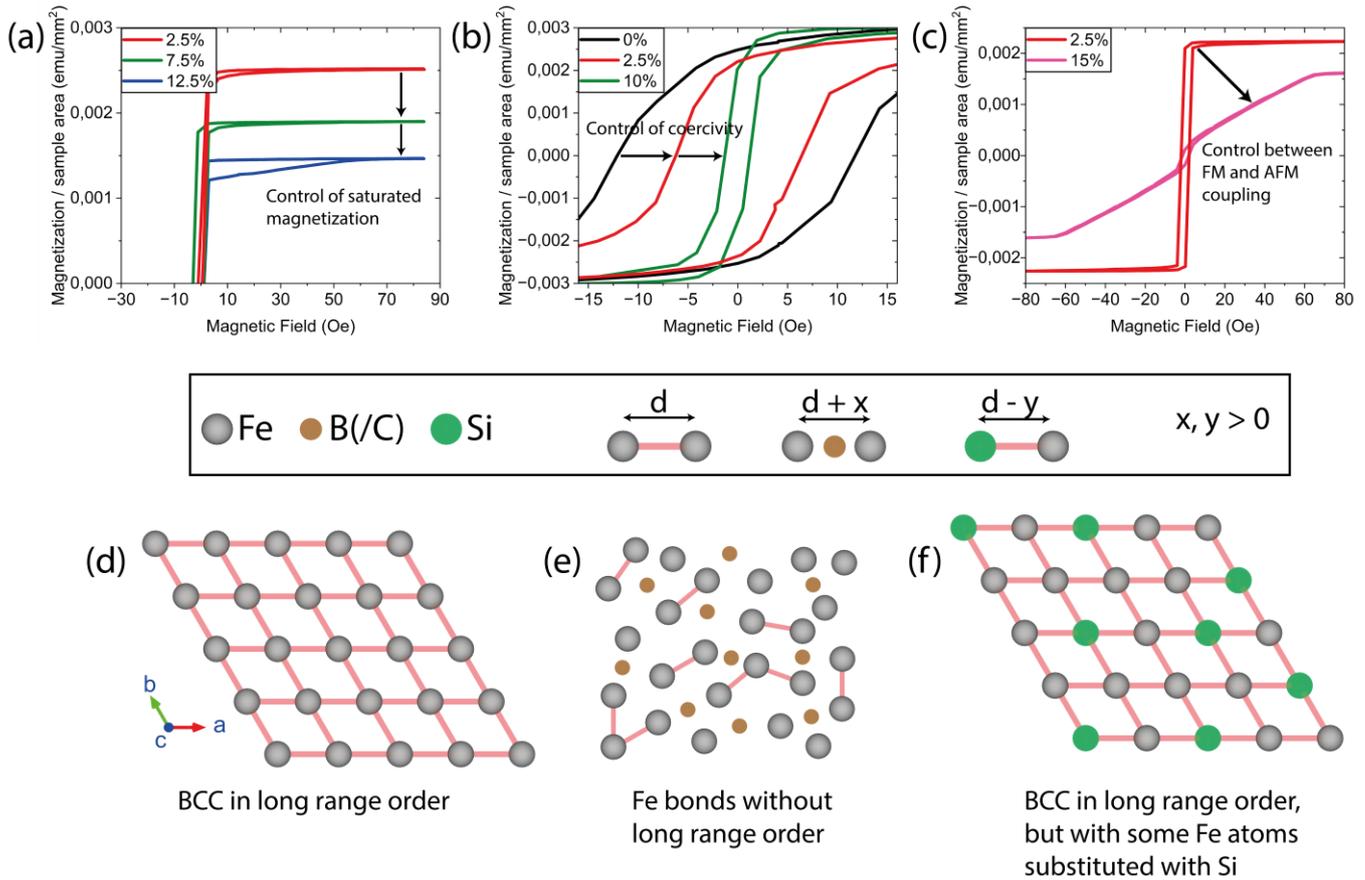

**Figure 6. Interpretation of hysteresis control.** Schematics of hysteresis control using data (a-c) and crystal structure (d-f). (a) showcasing the control of saturated magnetization using $B_4C$. (b) depicting the control of coercivity with $B_4C$. (c) demonstrating the possibility of controlling between FM and AFM coupled magnetic layers using different concentrations of $B_4C$. (d) shows an Fe BCC crystal structure, while (e) shows when B (/C) atoms are positioned in the interstitials, thus hindering medium to long-range order, thus appearing amorphous without breaking any significant amount of Fe-Fe bonds. (c) depicts the crystal structure at interfaces between Fe and Si layers, where some Fe atoms according to XRD, could have been substituted by Si atoms, thus decreasing the lattice parameter slightly.

The temperature-dependent studies (Figure S2) reveal a transition from antiferromagnetically coupled to ferromagnetically coupled phases below specific critical temperatures ($T_{CR}$).[29] Samples with 10% and 12.5% compositions exhibit weaker antiferromagnetic coupling compared to the 15% sample, as they transition to ferromagnetic coupling below 150K, whereas the 15% sample maintains partial antiferromagnetic coupling until 20K. The 15% sample demonstrates the strongest antiferromagnetic coupling, indicated by its lowest $T_{CR}$ and the largest field dependence in its M(T) curve, while the 2.5% sample is ferromagnetically coupled throughout.

Regarding the magnetic TEM, the complexity of Fe/Si multilayers amorphized by B$_4$C highlights the role of disrupted crystallinity and weakened interlayer interactions in creating weakened magnetic domain behaviours within different layers. Collectively, these results demonstrate how amorphization impacts the coherence and distribution of magnetic domains, while still retaining in-plane closure domain structures across all samples. The lack of thin and well-defined domain walls renders it easy to manipulate the in-plane magnetization.[30]

Conclusively, hysteresis loops can therefore be controlled in terms of saturated magnetization, coercivity and control between FM and AFM coupling, all depending on the concentrations of B$_4$C, as seen in Figure 6(a-c).

**The incorporation of B$_4$C and its' structural effects on the lattices**

Concerning the diffraction peak shift in the XRD pattern [Figure 3(e)], particularly if we compare the diffraction peak for the 50 Å sample with that for the 25 Å sample, is due to a decrease in Fe purity in the Fe layers and an increase in the amount of Fe with Si atoms in the lattices.[27] Since the 25 Å sample has twice as many interfaces than the 50 Å sample, there is more Fe + Si regions and considering the very thin layers possibly having Si atoms throughout the entirety of the Fe layers, thus the entirety of the sample, in the 25 Å periodicity samples. When the Si atoms substitute into the Fe positions in the layers lattices [Figure 6(f)], the diffraction peak will shift to higher angles according to Vegard's law.[31] This is because Si atoms are smaller than Fe atoms (Si = 1.11Å, Fe = 1.26 Å). The higher intensity is due to the overlapping in Fe/Si with Λ = 25 Å being a mix of larger sizes of coherent lattices of Fe + Si thus no pure Fe layers.

The X-ray diffraction (XRD) results reveal that the incorporation of B$_4$C into Fe layers leads to a progressive decrease in crystallite size as the B$_4$C concentration increases. Eventually, the sample transitions to a fully amorphous state. This transition is accompanied by a small shift in the diffraction peak to lower angles, which is attributed to the expansion of the lattice caused by B atoms occupying interstitial positions [Figure 6(d-e)].[32] The interstitial incorporation of B results in a slight lattice expansion, thereby shifting the XRD peak to lower 2θ values. However, due to the disruption of medium- to long-range crystallinity, the structure appears fully amorphous in the XRD analysis.

Interestingly, despite the apparent amorphous nature, X-ray photoelectron spectroscopy (XPS) and near-edge X-ray absorption fine structure (NEXAFS) analyses show that the amount of Fe-Fe bonds remains nearly unchanged.[20] This indicates the presence of short-range Fe-Fe lattices, even in samples that appear amorphous in XRD. Magnetic measurements further corroborate this finding, as all samples, including the amorphous ones, exhibit high magnetization. The observed ferromagnetic response is strong evidence for the persistence of Fe-Fe interactions, which are essential for sustaining magnetization.[33] Moreover, the saturation magnetization across the sample series remains nearly constant for most period thicknesses, regardless of the B$_4$C concentration. This consistency highlights the unique impact of B atoms: they disrupt long-range order without significantly affecting short-range Fe-Fe interactions or the material's magnetic properties. This distinctive behaviour of B in Fe layers underscores its potential as an effective amorphization agent that retains the ferromagnetic character of Fe.

# CONCLUSIONS

This study investigated the ability to control the magnetization amplitude, the magnetic coercivity, and the ferro-/antiferromagnetic coupling, in Fe/Si multilayers, by simply incorporating B$_4$C throughout the layers in different amounts. The results should withstand regardless of ferromagnetic layer and/or non-magnetic spacer layer material since it is the amorphization of the ferromagnetic layers by using B$_4$C that is crucial. The magnetic tuneability possibilities could give opportunities for applications within spintronics, magnetoresistance, data storage and sensors where many prevalent limitations are caused by the conditions of the magnetic hysteresis, by simply incorporating B$_4$C in magnetic multilayers. The XRD analysis reveals that incorporating B$_4$C into Fe layers causes lattice expansion until amorphization, evidenced by a shift in diffraction peaks to lower angles. Despite the disruption of medium- to long-range crystallinity, Fe-Fe bonds persist, as confirmed by XPS, NEXAFS and magnetic measurements such as magnetometry, polarized neutron reflectivity and microscopy. This unique behaviour maintains strong ferromagnetic properties even in the amorphous state, highlighting B$_4$C's potential as an effective amorphization agent while causing negligeable loss in magnetization.

## EXPERIMENTAL SETUP

Fe/Si and Fe/Si + $^{11}$B$_4$C multilayer thin films were deposited via ion-assisted magnetron sputter deposition in a high vacuum system,[34] maintaining a background pressure of approximately $5.6 \cdot 10^{-5}$ Pa ($4.2 \cdot 10^{-7}$ Torr). The films were layered onto 001-oriented single crystalline Si substrates, each measuring 10×10×1 mm³, with an existing native oxide layer. The substrate temperature was kept at room temperature (293 K), and to enhance thickness uniformity, the substrate rotated at 8 rpm. The substrate table was electrically isolated, allowing for the application of a substrate bias voltage to attract sputter gas ions from the plasma. A distinctive aspect of growing these samples, involved a modulated ion assistance approach during deposition. This process utilized ion-assisted deposition by attracting Ar-ions from the sputter plasma with a negative substrate bias of -30 V. A magnetic field aligned with the substrate normal helped concentrate the plasma toward the substrate, increasing the Ar-ion flux. The sputtering cycle began with a 0 V substrate bias for the initial approximately 3 Å, then switched to -30 V for the remainder of the layer to reduce intermixing. The sputtering targets comprised Fe (99.95% purity, 75 mm diameter), $^{11}$B$_4$C (99.8% chemical purity and >90% isotopic purity, 50 mm diameter), and Si (99.95% purity, 75 mm diameter). The magnetrons operated continuously during deposition, with material fluxes regulated by computer-controlled shutters for each target. This setup enabled the deposition of distinct multilayers and the co-sputtering of two target materials to achieve specific compositions. For the Fe/Si + $^{11}$B$_4$C films, each bilayer consisted of co-sputtered $^{11}$B$_4$C and Fe, followed by $^{11}$B$_4$C with Si. The deposition rates for Fe + $^{11}$B$_4$C and Si + $^{11}$B$_4$C were nearly identical, around 0.5 Å/s.

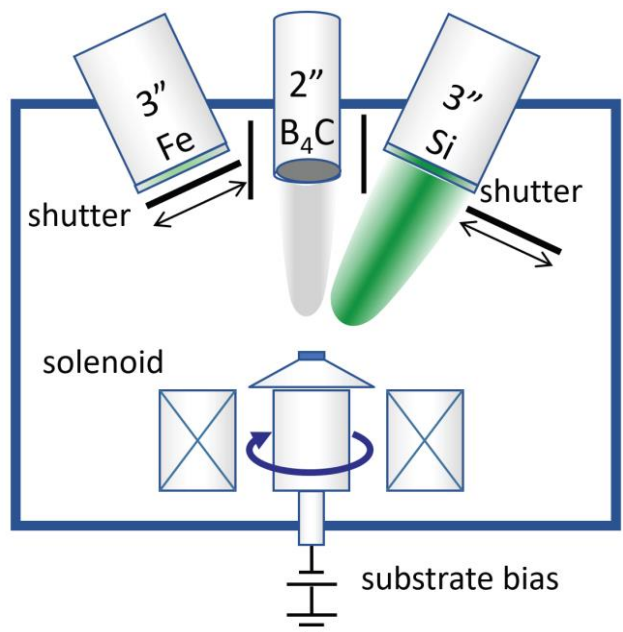

**Figure 7. Deposition system.** Schematic of the deposition system, allowing for co-sputtering of different targets simultaneously and computer-controlled shutters for layering.

Hard X-ray reflectivity analysis was conducted using a Panalytical Empyrean diffractometer. This setup involved a parallel beam configuration with a line-focused copper anode source, which operated at 45 kV and 40 mA to produce Cu-Kα radiation at a wavelength of 1.54 Å. A parabolic X-ray mirror and a ½° divergence slit were incorporated into the incident beam path to refine the beam and control the X-ray spot size on the sample. For the diffracted beam, a parallel plate collimator equipped with a 0.27° collimator slit was used, followed by a PIXcel detector in open detector mode.

X-ray diffraction (XRD) utilized a Panalytical X'Pert diffractometer in Bragg-Brentano geometry. This included Bragg-Brentano HD incident beam optics with a ½° divergence slit and a ½° anti-scatter slit. Secondary optics featured a 5 mm anti-scatter slit and an X'celerator detector in scanning line mode. Measurements were conducted over a 2θ range of 20° to 90°, with a step size of 0.018° and a collection time of 20 s.

For X-ray reflectivity (XRR) data analysis, the thickness of multilayer bilayers and interface roughness were evaluated through fitting procedures using the GenX software (version 3).[35] This software also facilitated the fitting of data obtained from X-ray reflectivity measurements.

Magnetic properties were characterized at room temperature using a vibrating sample magnetometer (VSM) in longitudinal geometry. Measurements were conducted over a field range of -1 T to 1 T to determine the saturation magnetization and saturation field. The VSM measurements were performed using the Vibrating Sample Magnetometer option of the Physical Property Measurement System (PPMS EverCool II, Quantum Design Japan, Inc.) at the CROSS User Laboratory II in Japan. Some of the temperature dependent measurements were measured in a SQUID magnetometer (MPMS2) from Quantum Design Japan, Inc.

XPS and NEXAFS measurements were conducted at the German-Eastern European Laboratory for Energy Materials Research Dipole beamline (GELEM Dipole, BESSY II, Berlin, Germany) operated by Helmholtz-Zentrum Berlin für Materialien und Energie.[36] All samples were mounted on a flag-style (Omicron) holder using conductive carbon tape. All measurements were performed at room temperature with the pressure in the analysis chamber of ca. $10^{-10}$ mbar. The photon energy was calibrated with Au4f ($E_{bin\_4f_{7/2}}$=84.00) in Au crystal. The NEXAFS spectrum was collected with total electron yield mode by sample drain current and normalized by beam current. Moreover, each sample was checked with five different points to make data reproducible. The fitting of XPS data was carried out with CasaXPS software.

The polarized neutron reflectivity (PNR) experiments were performed using the MORPHEUS polarized neutron reflectometer at the Paul Scherrer Institut in Villigen, Switzerland. Initially, a non-polarized neutron beam is polarized by a polarizer that filters the beam to permit only neutrons with a specific spin orientation. The spin state, either up or down, can be altered using a spin flipper. The beam, now polarized, is then aimed at a slight angle (θ) towards the sample, resulting in specular reflection. This process captures the reflected intensity, which is influenced by the reflective properties at each interface within the sample. PNR is particularly useful for assessing the spin-dependent scattering length density (SLD) of a sample, which provides insights into its magnetic characteristics. Two different spin orientations yield two separate PNR reflectivity profiles. Bragg peaks observed during the experiments indicate constructive interference. Measurements were conducted under an external magnetic field strong enough to magnetically saturate the samples in the in-plane direction, approximately 20 mT at the sample location, across a 2θ range from 0 to 15°, utilizing a neutron wavelength of 4.83 Å.

Thin-foiled specimens used in transmission electron microscopy (TEM) were prepared by using a system of focused ion beam (FIB; Helios Nanolab 650, FEI). Electron holograms, which produce the phase images representing the magnetic flux lines, and electron diffraction patters were collected with a 300 kV transmission electron microscope (HF-3300X, Hitachi). For the magnetic domain observations with electron holography, the thin-foil specimens were placed in a position free from the magnetic field generated by the objective lens. The electron holograms and electron diffraction patterns were recorded using a high-sensitivity camera (K3 camera, Gatan).

## AKNOWLEDGEMENTS

Hans Werthéns grant, 2022-D-03 (A.Z), Royal Academy of Sciences Physics grant, PH2022-0029 (A.Z), the Lars Hiertas Minne foundation grant FO2022-0273 (A.Z), the Längmannska Kulturfonden grant BA23-1664 (A.Z). J.S. is supported by the Japan Society for the Promotion Science (JSPS) KAKENHI Grants No. JP23H01840. We also acknowledge the Neutron Science and Technology Center, CROSS for the use of PPMS (VSM) in their user laboratories. We thank the Helmholtz-Zentrum Berlin für Materialien und Energie for the allocation of synchrotron radiation beamtime at GELEM Dipole. GELEM-PES Endstation at HZB received funding from the BMBF program ErUM-Pro. We thank the MORPHEUS beamline at PSI for the neutron beamtime.